\documentstyle[multicol,aps,amsfonts,amssymb,psfig,prl]{revtex}

\begin{document}
\draft
\preprint{{\bf ETH-TH/98-??}}

\title{\centerline{Quiet SDS' Josephson Junctions for Quantum
Computing}}

\author{Lev B.\ Ioffe$^{\ast,\ddagger}$, 
Vadim B.\ Geshkenbein$^{\dagger,\ddagger}$, 
Mikhail V.\ Feigel'man$^\ddagger$,
Alban L.\ Fauch\`ere$^\dagger$, 
and Gianni Blatter$^\dagger$}

\address{$^\ast$Department of Physics and Astronomy, Rutgers University,
Piscataway, NJ 08854, USA}
\address{$^\dagger$Theoretische Physik, ETH-H\"onggerberg, CH-8093 Z\"urich, 
Switzerland} 
\address{$^\ddagger$Landau Institute for Theoretical Physics, 117940 Moscow, 
Russia}
\date{January 11, 1999}
\maketitle

\begin{abstract}
\begin{center}
\parbox{14cm}
{
Unconventional superconductors exhibit an order parameter symmetry 
lower than the symmetry of the underlying crystal lattice. Recent 
phase sensitive experiments on YBa$_2$Cu$_3$O$_7$ single crystals 
have established the $d$-wave nature  of the cuprate materials, 
thus identifying unambiguously the first unconventional 
superconductor\cite{vanHarlingen,Tsuei}. The sign change 
in the order parameter can be exploited to construct a new 
type of $s$-wave--$d$-wave--$s$-wave Josephson junction exhibiting 
a degenerate ground state and a double-periodic current--phase 
characteristic. Here we discuss how to make use of these special 
junction characteristics in the construction of a quantum computer. 
Combining such junctions together with a usual $s$-wave link into 
a SQUID loop we obtain what we call a `quiet' qubit --- a solid 
state implementation of a quantum bit which remains optimally 
isolated from its environment.           
}

\end{center}

\end{abstract}

\pacs{PACS numbers: 85.25.Cp, 85.25.Hv, 73.23.-b, 89.80.+h}
\vspace{-0.4truecm}

\begin{multicols}{2}

Quantum computers take advantage of the inherent parallelism of the
quantum state propagation, allowing them to outperform classical
computers in a qualitative manner. Although the concept of quantum
computation has been introduced quite a while ago \cite{Feynman},
wide spread interest has developed only recently when specific
algorithms exploiting the character of coherent state propagation 
have been proposed \cite{Ekert}. Here we deal with the
device aspect of quantum computers, which is florishing in the wake of
the recent successes achieved on the algorithmic side.  Two
conflicting difficulties have to be faced by all hardware
implementations of quantum computation: while the computer
must be scalable and controllable, the device should be almost 
completely detached from the environment during operation 
in order to minimize phase decoherence. The most advanced 
propositions are based on trapped ions \cite{Cirac-Zoller,Monroe}, 
photons in cavities \cite{Turchette}, NMR spectroscopy of
molecules \cite{Gershenfeld}, and various solid state implementations
based on electrons trapped in quantum dots \cite{Loss}, the Coulomb 
blockade in superconducting junction arrays \cite{Schoen,Averin}, 
or the flux dynamics in Superconducting Quantum Interference Devices 
(SQUIDs) \cite{Bosco}. Nano\-structured solid state quantum gates 
offer the attractive feature of large scale integrability, once 
the limitations due to decoherence can be overcome \cite{Haroche}.

Here we propose a new device concept for a (quantum) logic gate
exploiting the unusual symmetry properties of unconventional
superconductors. The basic idea is sketched in Fig.\ 1: 
connecting the positive (100) and negative (010) lobes of a 
$d$-wave superconductor with a $s$-wave material produces 
the famous $\pi$-loop with a current carrying ground state
characteristic of $d$-wave symmetry \cite{vanHarlingen}.
Here we make use of an alternative geometry and match the $s$-wave
superconductors (S) to the (110) boundaries of the $d$-wave (D)
material. As a consequence, the usual Josephson coupling
$\propto (1 - \cos \phi)$ vanishes due to symmetry reasons and we arrive
at a bistable device, where the leading term in the coupling takes 
the form $E_d \cos 2\phi$ with minima at $\phi = \pm \pi/2$ (here,
$\phi$ denotes the gauge invariant phase drop across the junction). 
In our design we need the minima at the positions $\phi = 0,\pi$ --- 
the necessary shift is achieved by going over to an asymmetric SDS' 
junction with a large DS' coupling, see Fig.\ 1. The static DS' 
junction shifts the minima of the active SD junction by the desired 
amount $\phi = \pm \pi/2$. A similar double-periodic junction has 
recently been realized by combining two $d$-wave superconductors 
oriented at a $45^\circ$ angle\cite{Ilichev}.
\begin{figure} [bt]
\centerline{\psfig{file=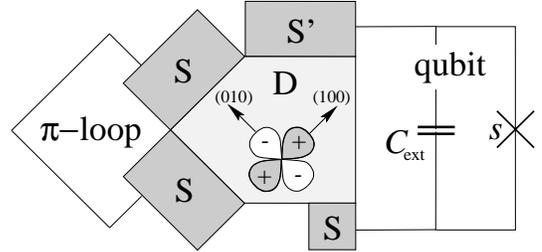,width=7cm,height=3.36cm}}
\narrowtext\vspace{4mm}
\caption{Geometrical arrangements between $s$-wave and $d$-wave
superconductors producing a $\pi$-loop (used in the phase sensitive
experiment by Wollman {\it et al.} [1]) and a qubit, the basic
building block of a quantum computer.}
\end{figure}
The ground states of our SDS' junction are degenerate and carry 
no current, while still being distinguishable from one another: 
e.g., connecting the junction to a large inductance loop, 
the $\pi$ state is easily identified through the induced current. 
It is this double-periodicity and the associated degeneracy in 
the ground state of the SDS' junction which we want to exploit 
here for quantum computation: combining the SDS' junction, 
a capacitor, and a conventional $s$-wave junction into a SDS' 
SQUID loop, we construct a bistable element which satisfies 
all the requirements for a qubit, the basic building block of 
a quantum computer. Below we give a detailed account of the 
operational features of our device.

Consider a small-inductance ($L$) SQUID loop with $I_{\rm
\scriptscriptstyle J}L\ll\Phi_0$, where $I_{\rm\scriptscriptstyle J}$ 
denotes the (Josephson) critical current of the loop and $\Phi_0 = 
hc/2e$ is the flux unit. Such a loop cannot trap magnetic flux 
($\Phi = 0$) and the gauge invariant phase differences $\phi_1$
and $\phi_2$ across the two junctions are slaved to each other,
as the uniqueness of the wave function requires that $\phi_1 - 
\phi_2 = 2\pi \Phi/\Phi_0$, see \cite{Tinkham}. Combining a SDS' 
junction with a coupling energy $E_d$ and a conventional 
$s$-wave junction (coupling $E_s$) into a SDS' SQUID loop, 
we obtain a potential energy
\begin{equation}
V (\phi) =  E_d (1-\cos 2 \phi) + E_s (1-\cos \phi),
\label{Pot-Energy}
\end{equation}
exhibiting two minima at $\phi = 0, \pi$, see Fig.\ 2. The switch 
$s$ allows us to manipulate their energy separation, choosing 
between minima which are either degenerate or separated by $2E_s$.

In the quantum case, the phase fluctuates as a consequence of the 
particle--phase duality \cite{Tinkham}. The phase fluctuations are
driven by the electrostatic energy required to move a Cooper pair
across the junction and are described by the kinetic energy
$T(\dot\phi) = (\hbar/2e)^2 C {\dot\phi}^2/2$, where $C$ denotes the 
loop capacitance. The dynamics of $\phi$ is manipulated by inserting 
a large switchable (switch $c$) capacitance $C_{\rm ext}$ into 
the loop acting in parallel with the capacitances $C_d$ and $C_s$ 
of the $d$- and $s$-wave junctions. Note that the Lagrangean 
$L = T - V$ of our loop is formally equivalent to that of a 
particle with `mass' $m \propto C$ moving in the potential $V(\phi)$. 

With the switch settings $c$ {\it on} and $s$ {\it off}, see Fig.\ 2(a),
the loop capacitance is large and the junction exhibits a doubly 
degenerate ground state which we characterize via the phase coordinate 
$\phi$, $|0\rangle$ and $|\pi\rangle$.
Closing the switch $s$, see Fig.\ 2(b), the degeneracy is lifted
and while $|0\rangle$ becomes the new ground state, the
$|\pi\rangle$-state is shifted upwards by the energy $2E_s$
of the $s$-wave junction, the latter being frustrated when
$\phi = \pi$. On the other hand, opening the switch $c$,
see Fig.\ 2(c), completely isolates the $d$-wave junction
and leads to the new ground and excited states $|\pm\rangle =
[|0\rangle \pm |\pi\rangle]/\sqrt{2}$ separated by the tunneling
gap $2\Delta_d$. The latter relates to the barrier $2E_d$ and
the capacitance $C_d$ of the $d$-wave junction via \cite{Tinkham}
$\Delta_d \propto E_d \exp(-2\sqrt{C_d E_d/e^2})$. Closing the 
switch $c$, the capacitance is increased by $C_{\rm ext}$ and 
the tunneling gap is exponentially suppressed. Using the above 
three settings, we can perform all the necessary single qubit 
operations:

{\it Idle-state:} The switch settings $c$-{\it on} and
$s$-{\it off} define the qubit's idle-state. While the large
capacitance $C_{\rm ext}$ inhibits tunneling, the degeneracy of
$|0\rangle$ and $|\pi\rangle$ guarantees a parallel time evolution
of the two states. This idle-state is superior to other
designs, where the two states of the qubit have {\it different}
energies and one has to keep track of the relative phase accumulated
between the basis states.

{\it Phase shifter:} Closing the switch $s$ separates the energies
of the basis states $|0\rangle$ and $|\pi\rangle$ by an amount
$2E_s$. Using a spinor notation for the two-level system, the relative
time evolution of the two states is described by the Hamiltonian
${\cal H}_s = -E_s \sigma_z$, with $\sigma_z$ a Pauli matrix.
Keeping the switch $s$ {\it on} during the time $t$, the time
evolution of the two states is given by the unitary rotation
$u_z(\varphi) = \exp (-i\sigma_z \varphi/2)$ with $\varphi =
-2 E_s t/\hbar$.
\begin{figure} [bt]
\centerline{\psfig{file=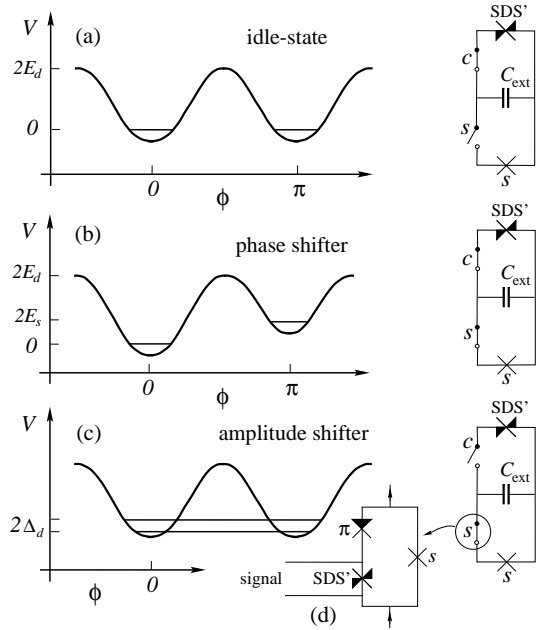,width=7cm,height=8.32cm}}
\narrowtext\vspace{4mm}
\caption{ Energy--phase diagrams for the SDS' SQUID loop.
(a) Idle-state:
The switches are set to $c$-{\it on} and $s$-{\it off} --- the
relative dynamics is quenched, leaving the state unchanged. 
(b) Phase-shifter:
With the switch settings $c$-{\it on} and $s$-{\it on} the 
relative phase between $|0\rangle$ and $|\pi\rangle$ increases
linearly with time.
(c) Amplitude-shifter:
The switch setting $c$-{\it off} isolates the $d$-wave junction. 
An initial state $|0\rangle$ oscillates back and forth between 
$|0\rangle$ and $|\pi\rangle$, allowing for a shift of amplitude.
(d) A SDS' junction, a $\pi$ junction, and a $s$-wave junction 
combined into a SQUID loop and serving as a switch.}
\end{figure}
{\it Amplitude shifter:} Assume we have prepared the loop in 
the ground state $|0\rangle$ and wish to produce a superposition 
by shifting some weight to the $|\pi\rangle$ state. Opening the 
switch $c$ in the loop, see Fig.\ 2(c), the time evolution
generated by the Hamiltonian ${\cal H}_d = \Delta_d \sigma_x$
of the open loop induces the rotation $u_x(\vartheta) = \exp 
(-i\sigma_x \vartheta/2)$ with $\vartheta = 2 \Delta_d t /\hbar$.
The system then oscillates back and forth between $|0\rangle$ 
and $|\pi\rangle$ with frequency $\omega = \Delta_d/\hbar$ 
and keeping the switch $c$  open for an appropriate time interval 
$t$ we obtain the desired shift in amplitude (note that the 
qubit remains isolated from the environment during these Rabi 
oscillations). 

Imposing the condition $E_d \gg E_s,~\Delta_d$ 
on the coupling energies, we make sure that the two states 
$|0\rangle$ and $|\pi\rangle$ are well defined while 
simultaneously involving only the low energy states 
$|0\rangle$ and $|\pi\rangle$ of the system. Furthermore,
all times involved should be smaller than the decoherence 
time $\tau_{\rm dec}$, requiring $E_s,~\Delta_d \gg \hbar/
\tau_{\rm dec}$.

The present setup differs significantly from the conventional 
(large inductance) SQUID loop design, where the low-lying states 
are distinguished via the different amount of trapped flux 
and their manipulation involves external magnetic fields $H$
or biasing currents $I$. SQUID loops of this type are being 
used in the design of classical Josephson junction computers 
\cite{Likharev} and have been proposed for the realization of
quantum computers, too, see \cite{Bosco}. However, this setup suffers
from the generic problem that the flux moving between the loops leads
to a magnetic field mediated long-ranged interaction between the
individual loops and further produces an unwanted coupling to the
environment. By contrast, our device remains decoupled from the 
environment, the operating states do not involve currents, 
and switching between states can be triggered with a minimal 
contact to the external world --- we therefore call our qubit 
implementation a `{\it quiet}' one.

Next, we discuss how to perform two-qubit operations within an array
of SDS' SQUID loops. A two-qubit state is a coherent superposition 
of single qubit states and can be expressed in the basis
$\{|xy\rangle\}$, where $x,y \in \{0,\pi\}$ denote the phases 
on the $d$-wave junctions of the first ($x$) and second ($y$) qubit, 
respectively. Unitary operations acting on these states are 
represented as $4\times 4$ unitary matrices. Single-qubit operations
$u$ acting on the second qubit take the block-matrix form
\begin{eqnarray}
{\cal U}_2 =
\left(\begin{array}{cc}\!
u & 0
\! \\ \!
0 & u
\! \end{array}\right),
\label{single_2}
\end{eqnarray}
and a similar block form selecting odd and even rows and columns 
defines the single-qubit operations on the first qubit. As all
logic operations on two qubits can be constructed from combinations
of single-qubit operations and the Controlled-NOT gate \cite{Ekert}
it is sufficient to define the operational realization of the latter.
The Controlled NOT gate performs the following action on two qubits: 
with the first (controller) qubit in state $|x\rangle$ and the
second (target qubit) in state $|y\rangle$ the operation shall
leave the target qubit unchanged if $x = 0$, while flipping it
between $0$ and $\pi$ when $x = \pi$, in matrix notation 
\begin{eqnarray}
{\cal U}_{\rm\scriptscriptstyle CNOT} =
\left(\begin{array}{cc}\!
{\rm{1\mskip-3mu I}} & 0
\! \\ \!
0 & \sigma_x 
\! \end{array}\right).
\label{CNOT}
\end{eqnarray}
The above Controlled NOT operation can easily be constructed
from the `phase shifter': Connecting two individual qubits in 
their idle-state over a $s$-wave junction into a SQUID loop, 
the states $|00\rangle$ and $|\pi\pi\rangle$ become separated 
from the states $|0\pi\rangle$ and $|\pi 0\rangle$ by the 
energy $2E_{s_b}$ of the $s$-wave junction. Keeping the two 
qubits connected during the time $t$ introduces a phase shift 
$\chi = -2E_{s_b}t/\hbar$ between the two pairs of states,
\begin{eqnarray}
{\cal U}_{\rm ps} (\chi) =
\left(\begin{array}{cc}\!
u_z(\chi) & 0 
\! \\ \!
0 & u_z(-\chi) 
\! \end{array}\right).
\label{phase_shifter}
\end{eqnarray}
The Controlled NOT gate (\ref{CNOT}) then can be constructed from the 
phase-shifter (\ref{phase_shifter}) via the following sequence of
single- and two-qubit operations (see \cite{Loss} for a similar 
realization of the CNOT gate),
\begin{eqnarray}
{\cal U}_{\rm\scriptscriptstyle CNOT} &=& \exp(-i\pi/4)
{\cal U}_{2y}(\pi/2){\cal U}_{1z}(-\pi/2){\cal U}_{2z}(-\pi/2)
\nonumber\\ &\quad\cdot& 
{\cal U}_{\rm ps}(\pi/2){\cal U}_{2y}(-\pi/2),
\label{reconstruct_CNOT}
\end{eqnarray}
where the single qubit operations ${\cal U}_{i\mu} (\theta)$ rotate 
the qubit $i$ by an angle $\theta$ around the axis 
$\mu$ ($u_\mu(\theta) = \exp(-i\sigma_\mu\theta/2)$ acting on $i$) while 
leaving the other qubit unchanged. 

A key element in our design are the switches and a valid suggestion
is the single electron transistor discussed in the literature \cite{Joyez}. 
Here we propose a quiet switch design optimally adapted to our 
SDS' qubits. The basic idea derives from frustrating the 
junctions in a SQUID loop resulting in a `phase blockade': 
Combining a SDS' junction with energy $E_d$, a $\pi$-junction 
with $E_\pi \ll E_d$, and a $s$-wave junction with $E_s = E_\pi$ 
into a (small inductance) SQUID loop, see Fig.\ 2(d), 
we obtain the following switching behavior: 
The phase $\phi = 0$ on the SDS' junction frustrates the 
remaining junctions and the loop's energy-phase relation 
is a constant, $E_{\rm sw} (\phi_\pi = \phi_s-\pi) \equiv 0$. 
A voltage pulse coming down the signal lines and switching 
the SDS' junction into the $|\pi\rangle$ state changes 
the phase relation between the $\pi$- and the $s$-wave
junctions and closes the switch: the energy $E_{\rm sw} 
(\phi_\pi = \phi_s) = 2 E_\pi (1-\cos\phi_\pi)$ produces the
current-phase relation $I = (2e/\hbar) \partial_{\phi_\pi} 
E_{\rm sw}$. The appropriate voltage pulses can be generated by 
driving an external SDS' SQUID loop unstable.

The quiet device concept proposed above heavily relies on the 
double periodicity of the SD junction. As the second harmonic is
strongly suppressed in a SID tunnel junction, a more feasible 
suggestion for the realization of a $\cos 2\phi$ junction is 
the SND sandwich, where the superconductors are separated by 
a thin metallic layer N. For a clean metallic layer, the coupling 
energies for the $n$-th harmonic are large and of order 
$E_{\rm\scriptscriptstyle J} \sim k_{\rm\scriptscriptstyle F}^2 
{\cal A}\, \hbar v_{\rm\scriptscriptstyle F}/d$, producing the 
well known saw-tooth shape in the current-phase relation 
\cite{Ishii} (here, $v_{\rm\scriptscriptstyle F}$ denotes the 
Fermi velocity in the N layer while $d$ and ${\cal A}$ are its 
width and area). In reality, it seems difficult to deposit 
a {\it clean} metallic film on top of a $d$-wave superconductor 
and we have to account for the reduction in the coupling 
$E_{\rm\scriptscriptstyle J}$ due the finite scattering length $l$ 
in the metal layer. Using quasi-classical techniques to describe 
a dirty SN$_{\rm\scriptscriptstyle D}$D junction, we obtain a second
harmonic coupling energy $E_d \sim k_{\rm\scriptscriptstyle F}^2 
{\cal A} \, (\hbar v_{\rm\scriptscriptstyle F}/d) (l/d)^3 \sim 
(R_{\rm\scriptscriptstyle Q}/R)(l/d)E_{\rm\scriptscriptstyle T}$, 
where $l$ denotes the scattering length in the normal metal, 
$R_{\rm\scriptscriptstyle Q} = \hbar/e^2$ is the quantum resistance,
and $E_{\rm\scriptscriptstyle T} \sim (\hbar 
v_{\rm \scriptscriptstyle F}/d) (l/d)$ is the Thouless energy.

The second important device parameter is the tunneling gap $\Delta_d$,
which depends quite sensitively on the coupling to the environment.
The usual reduction in the tunneling probability produced by the
environment \cite{CaldeiraLeggett} is reduced if the system is 
effectively gapped at low energies \cite{AmbegaokarEckernSchon}.  
This is the case for our SNDN'S' junction where the low-energy
quasi-particle excitations in the metal are gapped over the 
Thouless energy $E_{\rm\scriptscriptstyle T}$ \cite{Golubov}.
The dynamics of the junction is only affected by the 
presence of virtual processes involving energies larger 
than $E_{\rm\scriptscriptstyle T}$, leading to a renormalized 
capacitance $C_{\rm ren}\sim\hbar/R E_{\rm\scriptscriptstyle T}$ 
(cf.\ \cite{AmbegaokarEckernSchon}) and resulting in the reduced
tunneling gap $\Delta_d \propto E_d \exp[-\nu 
(R_{\rm\scriptscriptstyle Q}/R) \sqrt{l/d}]$, with $\nu$ 
of order unity. Consistency requires that the tunneling 
process is `massive' and hence slow, $\hbar/\tau < E_{\rm
\scriptscriptstyle T}$.  With a tunneling time $\tau \sim S/E_d$
($S\sim \hbar (R_{\rm\scriptscriptstyle Q}/R) \sqrt{l/d} =$ 
tunneling action) we find that the constraint $\hbar/\tau
E_{\rm\scriptscriptstyle T} \sim \sqrt{l/d} < 1$ is satisfied. 
The condition $\Delta_d \ll E_d$ requires the tunneling gap
$\Delta_d$ to be small, but large enough in order to allow for
reasonable switching times, requiring $(R_{\rm\scriptscriptstyle Q}/R)
\sqrt{l/d}$ to be of order 10. With typical device dimensions $d \sim
1000$ \AA, $l \sim 10$ \AA, and $R/R_{\rm\scriptscriptstyle Q} \sim
(d/l)(1/{\cal A}k_{\rm\scriptscriptstyle F}^2) \sim 1/100$, this
condition can be realized. Finally, the operating temperature 
$T$ is limited by the constraint $S/\hbar > E_d/T$, guaranteeing 
that our device operates in the quantum regime, and the requirement 
$T < E_{\rm\scriptscriptstyle T}$ that thermal quasi-particle 
excitations be absent. The first condition takes the form 
$T\ll \hbar/\tau \sim \sqrt{l/d} \, E_{\rm\scriptscriptstyle T}$ 
and is the more stringent one. Using the above parameters 
and a typical value $v_{\rm\scriptscriptstyle F} \sim 10^8$ cm/s,
we obtain a Thouless energy $E_{\rm \scriptscriptstyle T} \sim 1$ K 
and hence $T \ll 0.1$ K.

In conclusion, we have discussed a novel device concept for logic
gates in superconducting computers. The SDS' SQUID loop realizes a
number of attractive features which are potentially relevant both in
classical Josephson computers based on RSFQ logics as well as in
superconducting quantum computers. The most obvious advantage over
previous designs is the quietness of the device: The SDS' SQUID loop is
a naturally bistable device and does not involve external bias
currents or magnetic fields. Second, the basic states of the loop 
do not involve currents or trapped flux, hence long-range 
interactions between various elements of the computer are eliminated.  
Third, the qubits do not accumulate phase differences during idle 
time. And forth, all operations can be carried out via simple 
switching processes.

We thank Alexei Kitaev, Daniel Loss, Andrew Millis, and Boris Spivak for
discussions and acknowledge financial support from the Fonds National
Suisse.

\end{multicols}
\end{document}